\documentclass[
    ,final            
  ]
  {aipproc}

\layoutstyle{6x9}

\def\MET{{\mbox{$p\kern-0.42em\raise0.15ex\hbox{/}_{T}$}}}
\def\met{{\mbox{$p\kern-0.42em\raise0.15ex\hbox{/}_{T}$}}}
\def\mex{{\mbox{$p\kern-0.42em\raise0.15ex\hbox{/}_{x}$}}}
\def\mey{{\mbox{$p\kern-0.42em\raise0.15ex\hbox{/}_{y}$}}}

\begin{document}
\title{W/Z+jets measurements at D\O}

\classification{12.38.Bx, 13.85.Qk, 14.70.Fm, 14.70.Hp}
\keywords{Quantum Chromodynamics, W bosons, Z bosons}

\author{Darren D. Price\\(for the D\O\ Collaboration)}{
  address={Department of Physics, Indiana University, Bloomington, IN 47405, USA},
}
\begin{abstract}
We present a summary of recent measurements of vector boson production in association with jets in $p\bar{p}$
collisions at $\sqrt{s}=1.96$~TeV with the D\O\ detector. Results of measurements of the inclusive $n$-jet cross-sections, $\sigma_{n}/\sigma_{n-1}$ ratios 
of $W$+$(n)$jet production for $n=0-4$ and the normalized $n^\mathrm{th}$ jet $p_T$ differential cross-section distributions, and measurement of the 
production cross-section (times branching fraction) ratio of $Z+b$-jet to $Z$+jet production are reported. Measurement of the normalized 
$Z/\gamma^*$+jet angular cross-section differential distributions are also discussed.
\end{abstract}

\date{August 19, 2011}

\maketitle

\section{Introduction}

The distinctive signatures of $W/Z$+jets production allow for precision tests of perturbative QCD calculations
and are particularly important for high jet multiplicity 
events where next-to-leading order (NLO) predictions have recently become available. Measurements of these processes
also provide detailed benchmarks for the tuning of parton-shower (PS) and PS+matrix-element matched Monte Carlo (MC) generators.
$W/Z$+jets final states often dominate the event selections of many other Standard Model processes and new physics signals 
predicted at the Tevatron and the LHC, and these measurements can constrain these backgrounds, reducing uncertainties on their modeling.

\section{W+(n)jet cross-sections}

The production of jets in association with a $W$ boson was studied using data corresponding to an integrated luminosity of 4.2~fb$^{-1}$ collected with the D\O\ detector
in the semi-leptonic (electron) decay channel.
The following requirements were imposed on the event selection:
electron $p_T^e \ge 15$ GeV and pseudorapidity $|\eta^e| < $ 1.1, \met $>$ 20 GeV, $M_T^W \ge$ 40 GeV, jet $p_T \ge$ 20 GeV and $|y| < $ 3.2. 
Jets are reconstructed using the D\O\ RunII Midpoint Cone algorithm
with a cone size of ${\cal R}=0.5$.

The inclusive $W+(n)$jet cross-sections (for $n=0-4$) were measured in the above phase space, as was the $\sigma_{n}$ to $\sigma_{n-1}$ jet cross-section ratio (for $n=1-4$), 
where some experimental uncertainties cancel. In addition, the $W+(n)$jet cross-sections are measured differentially as a function of the 
$n^\mathrm{th}$ ($p_T$-ordered) jet $p_T$ in the $n^\mathrm{th}$ jet multiplicity bin.
The results are all fully corrected for the effect of finite experimental resolution, detector response, acceptance, and efficiencies back to the particle level, 
which includes energy from stable particles, the underlying event, muons, and neutrinos. Full details of the analysis and experimental results can be found in Ref.~\cite{wjets:2011}.
The unfolding procedure is performed using a Singular Value Decomposition (SVD) technique as implemented by the GURU~\cite{Hocker:1996} unfolding program.

\begin{figure}[tbp]
\includegraphics[scale=0.32]{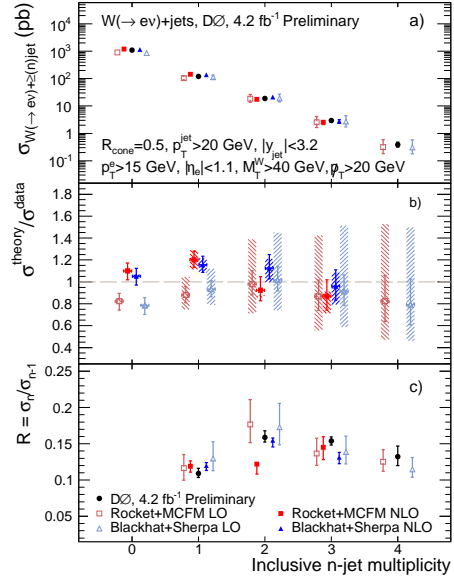}
\caption{\label{fig:inclXsec}
(a) Total inclusive $n$-jet cross sections $\sigma_n=\sigma (W(\to e\nu)\ + \geq \hspace{-1mm} n\textrm{ jet;}\; p^{\textrm{jet}}_T>20~\textrm{GeV})$
    as a function of inclusive jet multiplicity,
(b) the ratio of the theory predictions to the measurements, and
(c) $\sigma_n/\sigma_{n-1}$ ratios for data, \textsc{blackhat} and \textsc{rocket+mcfm}.
Error bars on data points represent combined statistical and systematic uncertainties on measured cross sections. The uncertainties on the theory points in (a) and (c)
and the hashed areas in (b) represent the theoretical uncertainty arising from the choice of renormalization and factorization scale. In (b) the error bars on the points
represent the data uncertainties.
}
\end{figure}

\begin{figure}
\centering
\begin{tabular}{cc}
\includegraphics[scale=0.32]{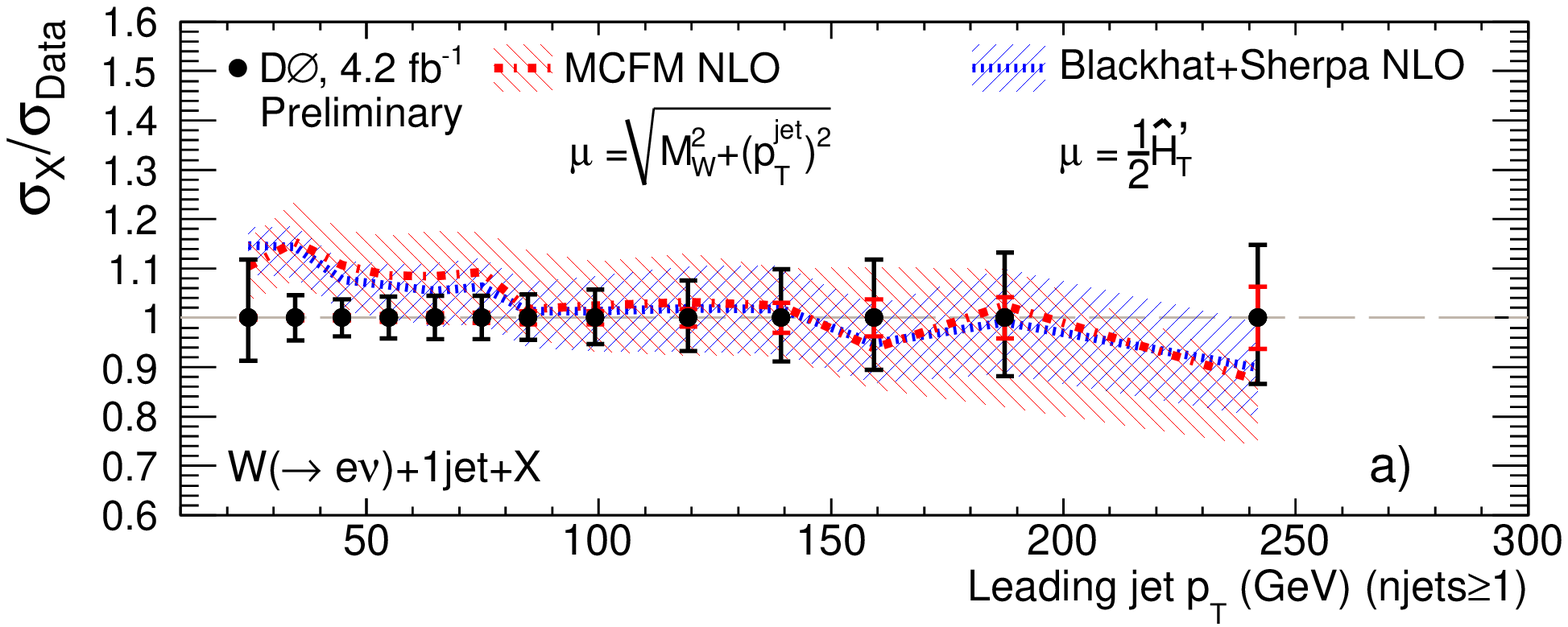} &
\includegraphics[scale=0.32]{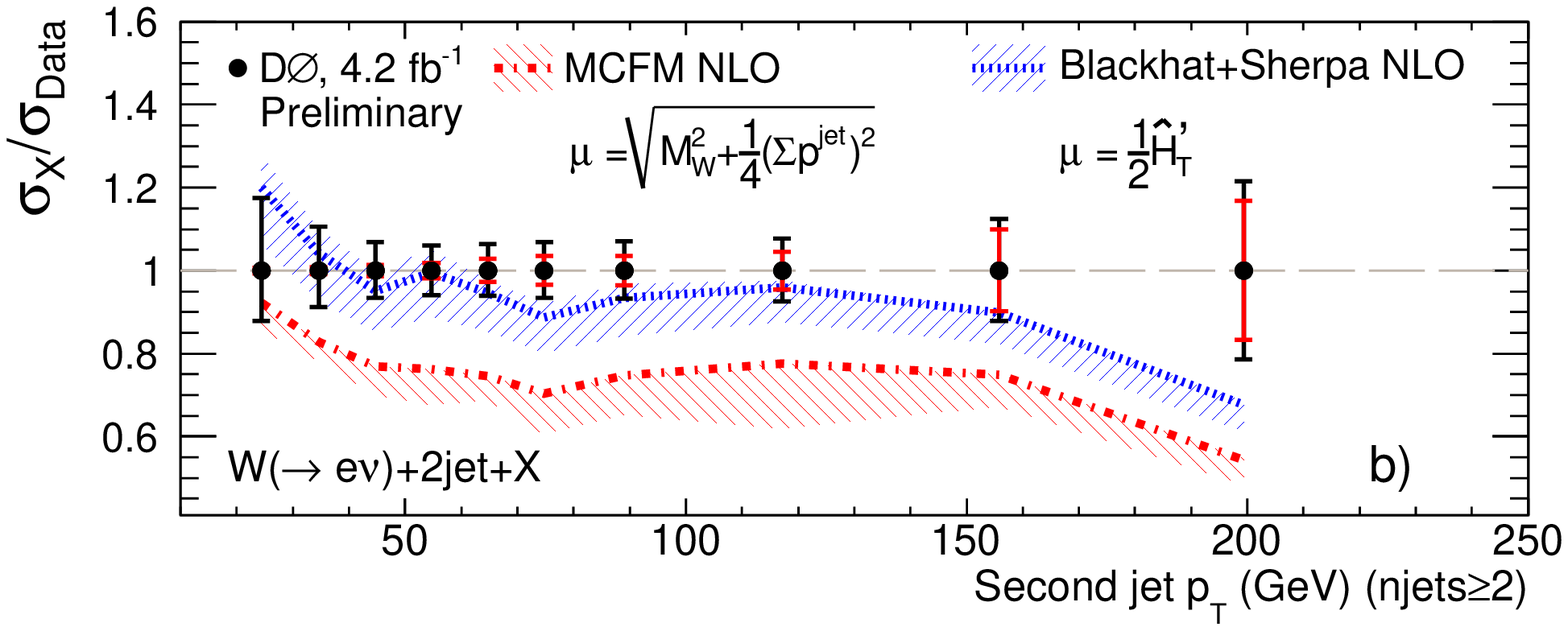}\\
\includegraphics[scale=0.32]{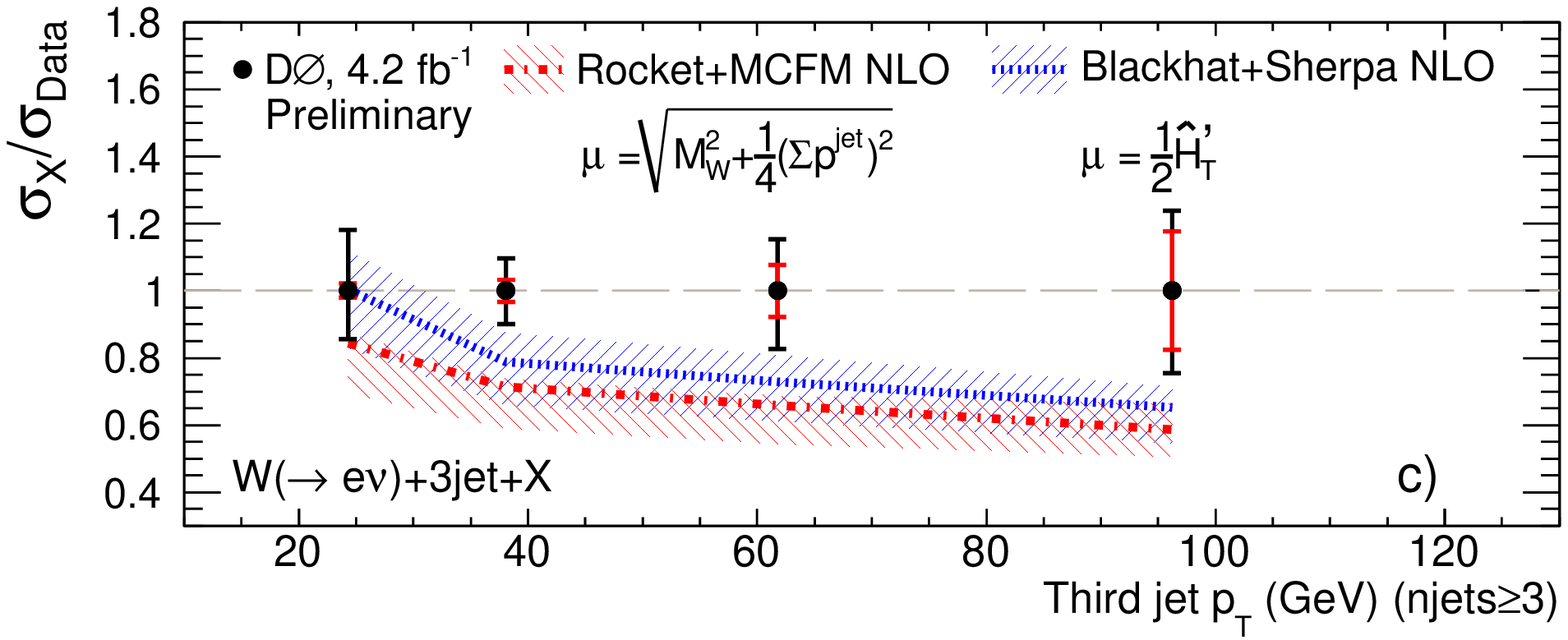}&
\includegraphics[scale=0.32]{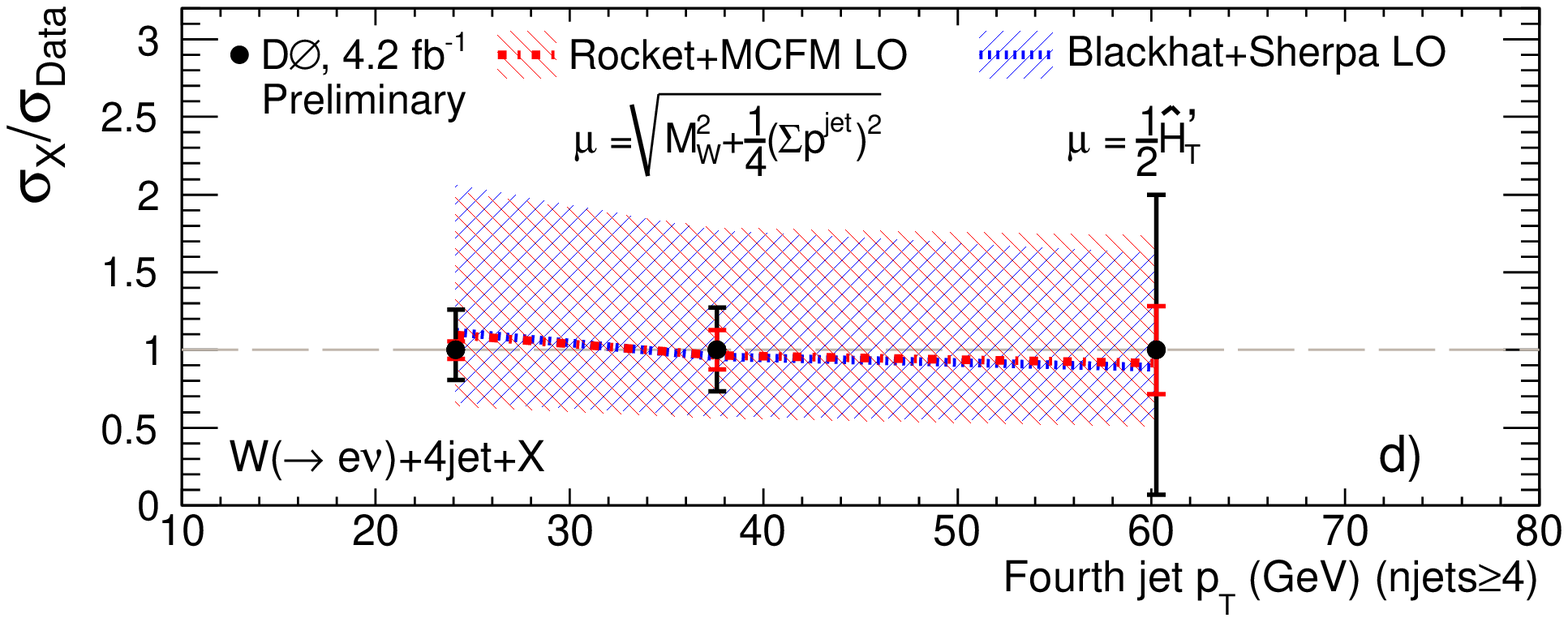}
\caption{\label{fig:ratios} The ratio of pQCD predictions to the measured differential cross sections for the $n^{\mathrm{th}}$ jet $p_T$ in
(a) $W$+1 jet events, (b) $W$+2 jet events, (c) $W$+3 jet events, and (d) $W$+4 jet events.
The data and theory predictions are normalized by the measured inclusive $W$ boson cross section and the predicted inclusive $W$ boson cross sections, respectively.
The inner (red) bars represent the statistical uncertainties of the measurement, while the outer (black) bars represent the statistical and systematic uncertainties
added in quadrature.  The shaded areas indicate the theoretical uncertainties due to variations of the factorization and renormalization scale. The central scale choice used
for each prediction is indicated in the figure.
}
\end{tabular}
\end{figure}

Figure~\ref{fig:inclXsec} shows the resultant unfolded total cross-section (times branching fraction) measurements of $W+(n)$jet production in the defined phase space in comparison with
two next-to-leading order (leading order for $W$+4jets) theoretical predictions using the MSTW2008 PDF,
and the $\sigma_{n}/\sigma_{n-1}$ ratios where some systematics cancel.
Fixed-order theoretical predictions provide results only at the parton-level so corrections for non-perturbative hadronization and underlying event effects 
are derived using the \textsc{sherpa v1.2.3} MC generator and applied to the theory in order to allow for comparison to measured data. 
Figure~\ref{fig:ratios} shows the corrected differential cross-sections of $W$+jet events, as a function of jet $p_T$ for each of the four inclusive jet multiplicity bins studied.
In order to reduce experimental uncertainties the measured differential cross-sections are normalized to the measured $W$ (0-jet) inclusive cross-section. 
Comparisons are made to (N)LO pQCD predictions\footnote{\textsc{MCFM(+Rocket)} predictions produced for these plots were produced with \textsc{MCFM v5.3}, 
which has recently been found have an issue affecting $W$+2jet calculations. This issue is corrected in \textsc{MCFM v6.0}, and results are found to be in closer agreement to predictions 
from \textsc{Blackhat}. See Ref.~\protect{\cite{wjets:2011}} for details.} (again corrected for non-perturbative effects using \textsc{sherpa})
and although on the whole theoretical predictions are found to be in agreement with data, areas are identified where some discrepancies are observed.

\section{Z+b/Z+jet inclusive cross-section fraction}

Measurement of the inclusive cross-section of $Z$ boson production with at least one $b$-quark jet to the inclusive $Z$+jets production cross-section 
was performed~\cite{zbratio:2010} with 4.2~fb$^{-1}$ of data collected with the D\O\ detector. Measuring the ratio to inclusive $Z$+jets allows for partial cancellation of
systematic uncertainties, providing a more precise measurement for comparison to theoretical predictions. Both the di-electron and di-muon decay channels 
were studied. Events were selected as follows: electron (muon) $p_T>$15 (10)~GeV, pseudorapidity $|\eta_e|<1.1$ OR $1.5<|\eta_e|<2.5$ ($|\eta_\mu|$<2),
jet $p_T \ge$ 20~GeV for the leading jet (and 15~GeV on any subsequent jets) and pseudorapidity $|\eta|<2.5$.
Events with missing transverse energy $>60$~GeV are rejected to suppress the background from $t\bar{t}$ production.

\begin{figure}[tbp]
\includegraphics[scale=0.32]{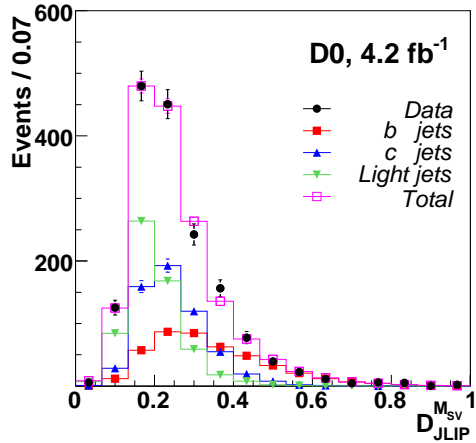}
\caption{\label{fig:zb} Distribution of the final neural net discriminant output for events in the combined lepton data sample, and the fitted light, $b$ and $c$ jet (and total) templates.
Uncertainties shown are statistical only.
}
\end{figure}

A challenge in this analysis is to extact the relatively small $Z+b$ signal from the overall selection. $Z$ boson candidates with a $b$-jet are separated from light and charm jet
candidates by using a neural network based b-tagging algorithm to distinguish the jet flavors. 
The final discriminant, $D^{M_SV}_\mathrm{JLIP}$, encoding all information from the neural network is calculated for each candidate event, and the resultant distribution is 
shown in Figure~\ref{fig:zb}. 
Templates are built as a function of this variable for light jets (determined from data by inverting the neural network selection requirements to
create a light-jet enriched data sample) and for charm/beauty jets (from MC simulation).
An unbinned maximum likelihood fit of these three templates to the data is performed for each of the di-lepton channels in order to extract the flavor fraction.
Consistent results are observed between both channels, so the data are combined and a new fit performed (see Figure~\ref{fig:zb}). 
The result of the combined maximum likelihood fit is
$\sigma(Z+b~\mathrm{jet})/\sigma(Z+\mathrm{jet})= 0.0193 \pm 0.0022~\mathrm{(stat)}\pm 0.0015~\mathrm{(syst)}$
which represents the most precise measurement of this ratio to date.
The largest systematics come from the discriminant template shape (4.2\%) and efficiency uncertainties (3.7\%).
An NLO \textsc{mcfm} prediction for the ratio yields $0.0185\pm 0.0022$ (corrected for hadronization and underlying event effects),
in reasonable agreement with the data measurement.

\section{Z+jets angular correlations}

Results described in Ref.~\cite{zangles:2009} present first measurement of angular correlations between the leading jet and the $Z/\gamma^*$
in $Z/\gamma^*(\to\mu^+\mu^-)$+jet events analyzed from 0.97~fb$^{-1}$ of data collected with the D\O\ detector.
The event selection requires $|\eta_\mu|<1.7$, $p_T(Z)>25~(45)$~GeV, jet  $p_T \ge$ 20~GeV and $|y| < $ 2.8.
Differential cross-sections are measured (normalized to the inclusive $Z$ boson cross-section, to reduce uncertainties) as a function of 
azimuthal angle, absolute rapidity difference, and the absolute value of the average rapidity of the $Z$ and leading jet.
These variables provide a unique test of pQCD calculations as they are sensitive to effects not probed in e.g. $p_T$ distributions.

Again, measurements are corrected back to particle-level accounting for detector resolution and efficiencies, and compared to 
NLO pQCD predictions (with \textsc{pythia}-derived non-perturbative corrections), as well as a selection of parton shower and PS+matrix element MC generators. 
Reasonable agreement is observed between NLO and data. Within MC generators studied, \textsc{sherpa} is found to best describe the shapes of the distributions, and areas
are observed where \textsc{alpgen}, \textsc{pythia} and \textsc{herwig} have problems in describing both shape and normalization of the data.

\vspace{-0.4cm}

\bibliographystyle{aipproc}   

\bibliography{ShortReferences}

\end{document}